%% file: manuscript.tex
\documentclass[manyauthors]{fundam}

\publyear{26}
\papernumber{3}
\volume{196}
\issue{1}
\theDOI{10.46298/fi.13347}

\versionForARXIV

\usepackage[english]{babel}
\usepackage{graphicx}
\usepackage{mathtools}
\usepackage{amsfonts}
\usepackage{amssymb}
\usepackage[autostyle]{csquotes}
\usepackage[font=small]{caption}
\usepackage[font=small]{subcaption}
\usepackage{hyperref}
\usepackage{siunitx}
\usepackage{chemformula}
\usepackage{orcidlink}

\newcommand{\codom}{\text{img}}
\newcommand{\dom}{\text{rng}}

\newcommand{\id}{\text{id}}
\newcommand{\mc}[1]{\mathcal{#1}}

\begin{document}

\title{Automated Inference of Graph Transformation Rules}

\address{\href{mailto:juri.kolcak@gmail.com}{juri.kolcak@gmail.com}}

\author{Jakob L. Andersen\,\affilref{1} \orcidlink{0000-0002-4165-3732}
\and
Akbar Davoodi\,\affilref{1} \orcidlink{0000-0001-6403-5091}
\and
Rolf Fagerberg\,\affilref{1} 
\and
Christoph Flamm\,\affilref{2} \orcidlink{0000-0001-5500-2415}
\and 
Walter Fontana\,\affilref{3} 
\and
Juri Kol\v{c}\'ak\,\affilref{1, 3, 4} \orcidlink{0000-0002-9407-9682}
\and
Christophe V.F.P. Laurent\,\affilref{1, 2}
\and
Daniel Merkle\,\affilref{1, 4} \orcidlink{0000-0001-7792-375X}
\and
Nikolai~N{\o}jgaard\,\affilref{1} 
}

\affils{
\affilref{1} Department of Mathematics and Computer Science, University of Southern Denmark; Odense, Denmark
\\ 
\affilref{2} Institute of Theoretical Chemistry, University of Vienna; Vienna, Austria
\\
\affilref{3} Department of Systems Biology, Harvard Medical School; Boston, Massachusetts, USA
\\
\affilref{4} Faculty of Technology, Bielefeld University; Bielefeld, Germany
}

\maketitle

\runninghead{Andersen \emph{et al.}}{Automated Inference of Graph Transformation Rules}

\begin{abstract}
The explosion of data available in life sciences is fueling an increasing demand for expressive models and computational methods. Graph transformation is a model for dynamic systems with a large variety of applications. We introduce a novel method of the graph transformation model construction, combining generative and dynamical viewpoints to give a fully automated data-driven model inference method.

The method takes as input dynamical properties, given as a \enquote{snapshot} of the dynamics encoded by explicit transitions, and constructs a compatible model. The obtained model is guaranteed to be minimal, thus framing the approach as model compression (from a set of transitions into a set of rules). The compression is permissive to a lossy case, where the constructed model is allowed to exhibit behavior outside of the input transitions, thus suggesting a completion of the input dynamics.

The task of graph transformation model inference is naturally highly challenging due to the combinatorics involved. We tackle the exponential explosion by proposing a heuristically minimal translation of the task into a well-established problem, set cover, for which highly optimized solutions exist. We further showcase how our results relate to Kolmogorov complexity expressed in terms of graph transformation.
\end{abstract}


\input{sections/introduction.tex}
\input{sections/preliminaries.tex}

\input{sections/methods.tex}

\input{sections/experiments.tex}
\input{sections/discussion.tex}

\section*{Acknowledgements}
This work was supported by the Novo Nordisk Foundation grants number NNF19OC0057834 and NNF21OC0066551; and by the Independent Research Fund Denmark, Natural Sciences, grant number DFF-0135-00420B.

\bibliographystyle{fundam}
\bibliography{references}
 
\end{document}

%% file: sections/introduction.tex

\section{Introduction}
\label{sec:intro}

Graphs are a very intuitive mathematical model of objects linked by relations. Being both visual and expressive, the versatility of graphs is underlined by their widespread use and adaptability. Graph transformation is a technique for specifying how one graph can be rewritten into another, thereby adding dynamics to the static modeling by graphs. Graph transformation is a powerful formalism well-founded in theory~\cite{EhrigPS73, HabelMP01, EhrigEHP04}. Graph transformation is not only a universal model of computation, but also enjoys a diverse and growing number of applications across numerous fields and areas, such as software engineering~\cite{HeckelT20, BrusELP87}, biology~\cite{Kniemeyer08, LoboVD11, BoutillierMLMKFCAF18, BlinovFGH04} or chemistry~\cite{AndersenFMS14, AndersenFMS16, AndersenFMS19, BehrK2021}.

In graph transformation, the rewriting of one graph into another is specified by the means of \emph{graph transformation rules}, or \emph{rules} for short. A rule consists of two graph patterns, one to match the input, and the other to specify the output. An application of a rule to a graph which contains a match of the input pattern replaces the matched part by the output pattern. A collection of such rules then defines a graph transformation model. Such a model encodes behavior in the usual form of a transition system, where states are graphs and transitions are all possible rule applications. This explicit representation of the encoded behavior may be arbitrarily larger than the rule set itself, possibly even infinite.

Numerous emerging applications of graph transformation, especially in the area of life sciences, challenge the traditional approach of first building a model (here, a set of rules) then analyzing the emergent behavior. Instead, we find ourselves in a situation where some, possibly all, of the transitions are known via empirical data, while the model which produces these transitions remains unknown. We are thus faced with the problem of reverse engineering a set of rules from their applications. We tackle this problem by introducing a fully automated method for construction of graph transformation models from (incomplete) transition systems.

The reverse engineering approach is especially relevant in the study of chemical reaction networks. Molecules are traditionally represented as undirected labeled graphs and chemical reactions can be naturally captured by the rules, or more precisely their applications. An empirically inferred chemical reaction network (e.g. metabolic network of a cell) can be considered the measurable expression of an unknown underlying model of the chemistry in question. Identifying this underlying model is a key question in network analysis. The simpler the model explaining the empirical data, the better, being a good rule of the thumb.

One should note that the graph patterns defining a rule might be arbitrarily large, including fully specified graphs. Each transition is thus, by itself, also a rule. Inferring a smaller rule set with the same semantics can be seen as an instance of a graph transformation model compression. Compressing a chemical reaction network, just as any other transition system, consists of identifying reactions which can be reproduced, or explained, by the application of the same rule. Chemically speaking, such reactions are likely to happen with the same, or at least highly similar underlying mechanism on the physicochemical level. As such, suggesting a model of chemistry by compression of known reactions is fundamentally of great interest.

The graph transformation model compression has further applications when the known chemical reaction network may be incomplete. Chemical reaction networks are often based on only partial empirical knowledge, especially in a biological setting, where the reactions themselves have to be reverse engineered from selected measurements. Metabolic networks pose a challenge in this aspect, due to the inherent complexity of the underlying chemistry. Often, a particular reaction is known for a single substrate only, but some similar molecules are very likely to be able to undergo the reaction in the same fashion. Such molecules are known as promiscuous substrates and are of interest in many areas of study, such as enzyme evolution, drug development or origin of life~\cite{AharoniGKGRT05, BornscheuerHKLMR12, CopleyNW23, Jensen76}.

We demonstrate that one can obtain suggestions for promiscuous substrates by performing a lossy compression of the graph transformation models. In particular, instead of demanding the rule set to reproduce exactly the input transitions, one can obtain an over-approximation by allowing for new rule applications, and thus transitions, in a controlled fashion. A lossy compression of a chemical reaction network thus results in a rule set which not only reproduces all the original reactions, but also suggests new reactions, which operate on the same underlying mechanisms (rules) as the existing ones, effectively performing a network completion. From a purely mathematical standpoint, one can also consider lossy compression to refer to a rule set that does not introduce any new transformations, but instead loses some of the original ones (under-approximation), or a combination of both. Due to lack of direct applications for the under-approximation case, however, we limit ourselves to lossy compression by over-approximation.

Last but not least, we can turn to the nature of the graph transformation model compression itself. In particular, by maximizing the compression ratio (minimizing the size of the rule set) we obtain a measure of complexity of graph transformation models in the minimal number of rules required to express the desired behavior. Naturally, such a complexity measure can be used to compare various empirical models, such as chemical reaction networks. From a computational perspective, the introduced complexity measure closely resembles Kolmogorov complexity. Indeed, we end up with an approximation of Kolmogorov complexity restated for graph transformation problems. A precise formulation requires additionally a consideration of the size of the rules themselves, not just their count.

In short, our contribution is a formal method for compressing the known transitions (explicitly represented semantics) of a graph transformation model into a set of rules (implicit semantics). The method has two main working modes, a lossless and a lossy compression, the latter producing an over-approximation of the input. We discuss various use-cases for the method, including reverse engineering, complexity analysis and model completion and present application examples across various graph transformation models.

The rest of the article is organized as follows. In Section~\ref{sec:prelim}, we recall the necessary graph transformation definitions and formalize the notion of transition systems as an explicit account of graph transformation model behavior. Section~\ref{sec:methods} describes in detail the method of graph transformation model compression as construction of a generating rule set. Some illustrative application examples are provided in Section~\ref{sec:experiments}. Finally, Section~\ref{sec:discussion} provides a summary of our contribution and some concluding remarks.
\subsubsection*{Notation}
Throughout the paper, we use the notation $B(A)$ to indicate the member $B$ of a tuple $A=(B, \dots)$.

At several points, we need to refer to the weakly connected components of graphs. We use $\oplus$ to denote the disjoint union on graphs and given a graph $G = g_1 \oplus \dots \oplus g_k$ with $k\in\mathbb{N}$ weakly connected components, we use $[G] = \{\!\{g_1, \dots, g_k\}\!\}$ to refer to the multiset of the connected components.

The paper contains several commuting diagrams for illustrative purposes. Different types of graph functions are represented by different arrow types. In particular:

{
    \includegraphics{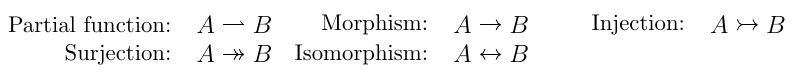}
}

%% file: sections/preliminaries.tex

\section{Preliminaries}
\label{sec:prelim}

In this section we establish the necessary formal foundations. First, we reiterate the relevant parts of double-pushout (DPO) framework for graph transformation. Second, we introduce the notions necessary for the method introduced in~Section~\ref{sec:methods} within the graph transformation framework, such as the transition system.

\begin{definition}[Graph]
    A graph $G = (V, E)$ is a couple consisting of a set of vertices $V = V(G)$ and a set of edges $E = E(G)$. An edge $e \in E(G)$ is a pair $({{}^\bullet}e, e^\bullet)$ of multisets of elements of $V(G)$.
    \label{def:graph}
\end{definition}

The above definition of a graph follows the intuition of \enquote{objects linked by relations} and is permissive enough to subsume most commonly employed graph types, including hyper-graphs.
Let now $\Sigma$ be a set of labels, or colors. Both vertices and edges may be assigned a label by the labeling function $\lambda\colon V(G) \cup E(G) \rightharpoonup \Sigma$. For ease of notation, we use simply $\lambda$ to denote the labeling function of any graph and we write $\lambda(x) = \epsilon$, where $\epsilon \notin \Sigma$ is a special empty label, for the elements $x \in V(G) \cup E(G)$ for which $\lambda$ is not defined.

To be able to proceed with DPO rule definition, we need to define graph morphisms. In the following we use $\dom(f)$ and $\codom(f)$ to refer to the range and image, respectively, as the specific subsets of the domain and codomain on which the function $f$ is defined.

A partial graph function $f\colon G \rightharpoonup H$ is a partial function on the vertex and edge sets of the graphs $G$ and $H$ (i.e. $\dom(f) \subseteq V(G) \cup E(G)$ and $\codom(f)\subseteq V(H) \cup E(H)$) fulfilling $f(V(G)\cap \dom(f)) \subseteq V(H)$ and for all edges $e\in E(G)\cap \dom(f)$ we have ${{}^\bullet}e \cup e^\bullet \subseteq \dom(f)$, $f(e) \in E(H)$, $f({{}^\bullet}e) = {{\vphantom{()}}^\bullet}f(e)$ and $f(e^\bullet) = {f(e)}^\bullet$. We say that a graph function is total, $f\colon G \rightarrow H$, if its defined on all the vertices of $G$, $V(G)\subseteq \dom(f)$.

A (partial) graph function $f\colon G \rightharpoonup H$ is a (partial) graph morphism if it satisfies both of the following properties:
\begin{itemize}
    \item[-] For all edges $e \in E(G)$ such that ${{}^\bullet}e \cup e^\bullet \subseteq \dom(f)$, $e \in \dom(f)$ (preserves edges).
    \item[-] For all $x \in \dom(f)$, $\lambda(x) \in \{\epsilon, \lambda(f(x))\}$ (preserves labels).
\end{itemize}

A (partial) graph morphism which is also injective, is called a (partial) graph monomorphism. If there exists a monomorphism $f\colon G \rightarrow H$ such that the inverse $f^{-1}\colon H\rightharpoonup G$ preserves labels, we say that $G$ is a subgraph of $H$, $G \subseteq H$.

A graph monomorphism $f\colon G\rightarrow H$ which is surjective, $V(H)\subseteq\codom(f)$, and whose inverse $f^{-1}\colon H \rightarrow G$ is a graph morphism is called a graph isomorphism. If there exists a graph isomorphism $f \colon G \rightarrow H$, we say that $G$ and $H$ are isomorphic, $G \simeq H$.

Finally, we lift the notion of isomorphism to the (partial) graph functions themselves. Two (partial) graph functions $f\colon G \rightharpoonup H$ and $f'\colon G \rightharpoonup H'$ with $\dom(f) = \dom(f')$ are isomorphic, $f \equiv f'$, if there exists a graph isomorphism $\varphi_H\colon H\rightarrow H'$ such that $\varphi_H \circ f = f'$.

\begin{definition}[Rule]
    A graph transformation rule is a span $p = (L \xleftarrow{l} K \xrightarrow{r} R)$, where $L, K$ and $R$ are graphs, called the left, invariant and right graph, respectively. $l\colon K\rightarrow L$ and $r\colon K\rightarrow R$ are graph morphisms, called the left and right morphisms, respectively, of which $l$ is additionally injective (a monomorphism).
    \label{def:rule}
\end{definition}

Mapping of the elements (vertices and edges) of $K$ into both $L$ and $R$ by virtue of $l$ and $r$, respectively, links them across the rule, from the input (left) graph to the output (right) graph and vice-versa. We refer to said link as the \emph{element map} $f_p\colon L\rightharpoonup R$, formally $f_p = r \circ l^{-1}$ with $\dom(f_p) = \codom(l)$ and $\codom(f_p) = \codom(r)$.

We are only interested in \emph{proper} rules, that is rules that only specify weakly connected components that are somehow modified from the left to the right side. Formally, we can use the element map to write $\forall g\in[L], g \neq f(g)$. Note that we demand inequality rather than non-existence of an isomorphism, $f$ constrained to $g$ could thus still be non-identity automorphism, $f_{/g} \not\equiv \id_g$.

\begin{figure}[!ht]
	\centering

    \includegraphics{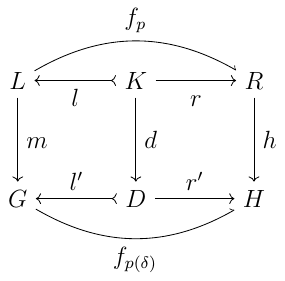}
    
	\caption{Commuting diagram of a rule $p = (L\xleftarrow{l}K\xrightarrow{r}R)$ applied to a graph $G$ using the match $m$, yielding a direct derivation $\delta = G \xRightarrow{m, p} H$. By definition, $p(\delta) = (G\xleftarrow{l'}D\xrightarrow{r'}H)$ is also a rule. The partial functions $f_p$ and $f_{p(\delta)}$ represent the element maps of the two rules.}
	\label{fig:derivation}
\end{figure}

An application of a rule $p = (L \xleftarrow{l} K \xrightarrow{r} R)$ to a graph $G$ consists of two phases. First, a graph morphism $m: L \rightarrow G$, called a \emph{match}, is selected. Second, what is in $L$ but not in $K$ is removed from $G$ (pushout complement of $K \xrightarrow{l} L \xrightarrow{m} G$), and finally, what is in $R$ but not in $K$ is added to produce the result $H$ (pushout of $D \xleftarrow{d} K \xrightarrow{r} R$). This process is formalized by the commuting diagram in Figure~\ref{fig:derivation}. The application of a rule $p$ to $G$ using a match $m$ is referred to as a \emph{direct derivation}, denoted by $\delta = G \xRightarrow{m, p} H$.

Several versions of DPO graph transformation, varying in expressiveness have been con\-side\-red~\cite{HabelMP01}, characterized by injectivity requirements on the right morphism and the match. Our definition of a rule and application represents the most general case, when no injectivity demands are made on either of the morphisms in question. For ease of presentation, however, we limit our examples to the case when both the match and right morphism are monomorphisms.

\begin{figure}[!ht]
	\centering
 
    \includegraphics{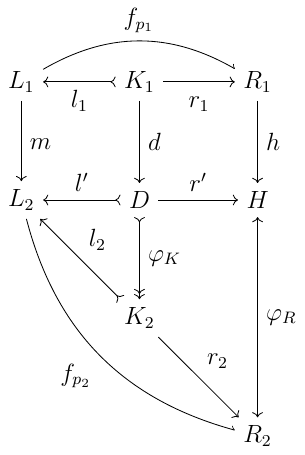}
 
	\caption{Commuting diagram of a subrule relation between rules $p_1 = (L_1\xleftarrow{l_1}K_1\xrightarrow{r_1}R_1)$ and $p_2 = (L_2\xleftarrow{l_2}K_2\xrightarrow{r_2}R_2)$.}
	\label{fig:subrule}
\end{figure}

Observe that a direct derivation $\delta$ encapsulates a span $G \xleftarrow{l'}D\xrightarrow{r'}H$, which is by itself a rule. We refer to said rule as \emph{induced} by the direct derivation, denoted by $p(\delta)$. The induced rule explicitly includes all the unchanging parts of $G$, expanding upon $p$ by including more invariant elements and thus being more specific (less applicable) than $p$. We generalize this property to any two rules $p_1$ and $p_2$ such that $p_2$ can be obtained as the induced rule of a direct derivation of $p_1$ and call $p_1$ a \emph{subrule} of $p_2$. This is illustrated by the commuting diagram in Figure~\ref{fig:subrule}.

\begin{definition}[Subrule]
    Let $p_1 = (L_1\xleftarrow{l_1}K_1\xrightarrow{r_1}R_1)$ and $p_2 = (L_2\xleftarrow{l_2}K_2\xrightarrow{r_2}R_2)$ be two rules. We say $p_1$ is a subrule of $p_2$, $p_1 \subseteq p_2$, if there exists a match $m\colon L_1 \rightarrow L_2$ such that $\delta = L_2 \xRightarrow{m, p_1} H$ is a direct derivation with $f_{p(\delta)} \equiv f_{p_2}$. The rules $p_1$ and $p_2$ are said to be isomorphic, $p_1 \simeq p_2$, if they are subrules of each other, $p_1 \subseteq p_2$ and $p_2 \subseteq p_1$.
\end{definition}

\eject

The subrule relation $p_1 \subseteq p_2$ ensures that whenever the rule $p_2$ is applied, $\delta_2 = G \xRightarrow{m_2, p_2} H_2$, there exists a direct derivation $\delta_1 = G \xRightarrow{m_1, p_1} H_1$ such that $f_{p(\delta_1)} \equiv f_{p(\delta_2)}$. It suffices to use the match $m\colon L_1\rightarrow L_2$ from the subrule definition in the construction of the match $m_1 = m_2 \circ m$. The rule isomorphism defined by the means of the subrule relation thus relates rules by their applications rather than their structure, as is the case for isomorphisms of the underlying spans. Any two isomorphic rules apply to the exact same graphs with the exact same results. It therefore suffices to consider the equivalence classes defined by the isomorphism, or more precisely, their canonical representatives\footnote{The equivalence classes can be infinite in general, but a total order minimizing the size of the graphs $L$, $K$ and $R$ and maximizing $\dom(\lambda)$, that is $\lambda(x) = \epsilon$ for an $x \in V(K) \cup E(K)$ only if $\lambda(l(x)) \neq \lambda(r(x))$, identifies a canonical representative.}. From now on, we identify all rules with their canonical representative.

The subrule relation is reflexive, as one can take the match $m = \id_L\colon L \rightarrow L$ to be the identity, and transitive, the match $m=m_2\circ m_1 \colon L_1 \rightarrow L_3$ being the composition of the subrule relation matches $m_1\colon L_1\rightarrow L_2$ and $m_2\colon L_2\rightarrow L_3$. The subrule relation lifted to the isomorphism classes (canonical representatives) is thus a partial order. Example~\ref{ex:subrules} illustrates the subrule relation on three rules, including a minimal subrule, as well as an example of rules which are not subrules of one another in spite of high structural similarity.

The ranges of the element maps $f_{p(\delta)}$ and $f_{p_2}$ from Figure~\ref{fig:subrule} are by definition equal to the images of $l'$ and $l_2$, respectively. The isomorphism $f_{p(\delta)} \equiv f_{p_2}$ thus gives us the equality $\codom(l') = \codom(l_2)$ and by extension, the function ${l_2}^{-1} \circ l'=\varphi_K\colon D \rightarrow K_2$ is not only injective, but also total and surjective. It is not necessarily a morphism, however, as it is not guaranteed to preserve labels. It is intuitive to see that $\varphi_K$ will be a morphism in case $p_2$ is the canonical representative of its subrule isomorphism class, due to its invariant graph $K_2$ specifying all possible labels.

\input{examples/subrules.tex}

What remains to be defined is the input structure for the compression method outlined in Section~\ref{sec:methods}. The (labeled) transition system formalism is commonly employed for representation of discrete model behavior. Intuitively, a transition system consists of states, graphs in our case, and transitions between them, direct derivations in our case. A direct derivation $\delta = G \xRightarrow{m, p} H$ carries more information than one may expect in an input transition system, that is when the model is unknown, namely the rule and the match. Instead, we consider the transitions to be equipped with the element map $f\colon G \rightharpoonup H$ only.

The above intuitive definition of a transition system indeed suffices to capture the graph transformation model dynamics. We consider a slightly modified version of the transition systems for our method input, however, as we do not require the transition system to be comprehensive, only to provide sufficiently large snapshot of the model behavior. This modification consists in delimiting a special set of weakly connected graphs we refer to as input graphs. Explicit specification of the input graphs is motivated by the need for a simple formalism to encompass the varied needs of our application scenarios. In particular, the graphs representing states in a chemical reaction network typically consist of several (weakly) connected components (molecules) and allow for arbitrary recombination for the purpose of rule application. E.g. a molecule $A$ may react with molecule $B$ and also with molecule $C$, but $B$ and $C$ may not react together. Having the connected components be the input graphs thus eliminates the need to construct all possible combinations explicitly. This is further illustrated in Example~\ref{ex:input_transition_system}.

\begin{definition}[Input Transition System]
    An input transition system is a tuple $S = (\mc{U}, \mc{T})$ where $\mc{U} = \mc{U}(S)$ is a finite set of weakly connected input graphs and $\mc{T} = \mc{T}(S)$ is a finite set of transitions of the form $t = (G, f, H) \in \mc{T}(S)$, such that $G$ and $H$ are graphs, of which $[G]\subseteq\mc{U}(S)$ and $f\colon G \rightharpoonup H$ is a partial graph function representing the element map. Similar to rules, we are only interested in proper transitions, $\forall g\in[G], g\neq f(g)$.
    
    We use the arrow notation $t = G \xrightharpoonup{f} H$ to denote a transition $t = (G,f,H) \in \mc{T}(S)$.
\end{definition}

\input{examples/input_transition_system.tex}

%% file: examples/subrules.tex

\begin{example}[Subrules]
\label{ex:subrules}

Consider the rule $p = (L \xleftarrow{l} K \xrightarrow{r} R)$ and three additional rules, $p_0 = (L_0 \xleftarrow{l_0} K_0 \xrightarrow{r_0} R_0)$, $p_1 = (L_1 \xleftarrow{l_1} K_1 \xrightarrow{r_1} R_1)$ and $p_2 = (L_2 \xleftarrow{l_2} K_2 \xrightarrow{r_2} R_2)$, each of which matches the left side of $L$, $m_0\colon L_0\rightarrow L$,$m_1\colon L_1\rightarrow L$ and $m_2\colon L_2\rightarrow L$, as depicted in Figure~\ref{fig:subrules}.

\begin{figure}[!ht]	
	\centering

    \includegraphics[scale=1.2]{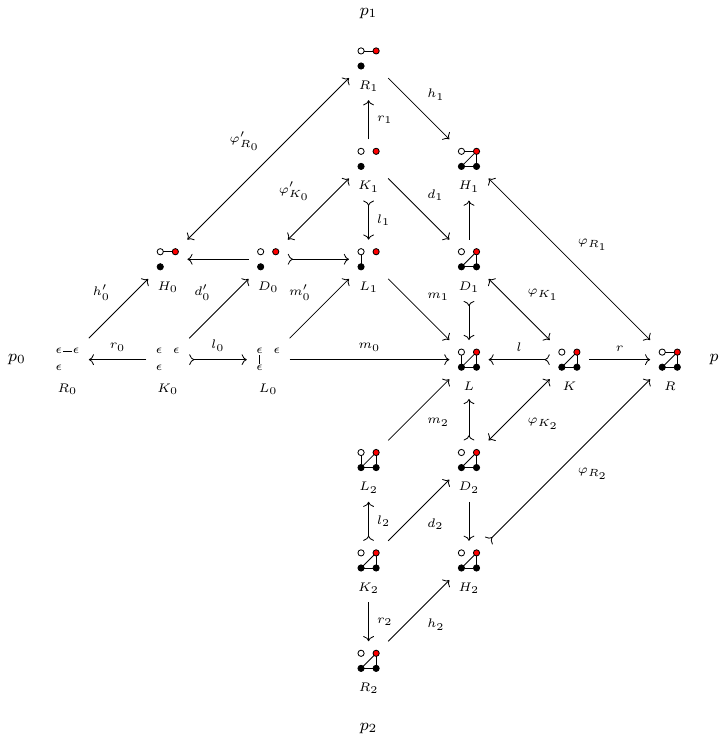}
	
	\caption{%
		 A commuting diagram showing a rule $p$ and three other rules, $p_0, p_1$ and $p_2$ with matches into the left side $L$. The relative position of the vertices is used to indicate the graph morphisms. $p_1$ is shown to be a subrule of $p$ and $p_0$ a subrule of both $p$ and $p_1$. $p_2$ is not a subrule of $p$ as $\varphi_{R_2}$ is just a monomorphism, not an isomorphism.%
	}
	\label{fig:subrules}
\end{figure}

The rule $p_1$ is easily shown to be a subrule of $p$ as the application to $L$ according to $m_1$ yields the graphs $D_1$ and $H_1$, which are identical, and thus trivially isomorphic to $K$ and $R$. Similarly, $p_0$ is shown to be a subrule of $p_1$ using a match $m_0'\colon L_0\rightarrow L_1$ such that $m_0 = m_1 \circ m_0'$. Rules $p_0$ and $p_1$ also illustrate the transitivity of the subrule relation. The application of $p_0$ to $L$ by $m_0$ can be reconstructed by taking $d_0 = d_1 \circ \varphi_{K_0}' \circ d_0'$ and $h_0 = h_1 \circ \varphi_{R_0}' \circ h_0'$. Indeed, $p_0$ is in fact a minimal subrule of $p$.

The rule $p_2$, on the other hand, is not a subrule of $p$. Using the only possible match $\id_{L} \equiv m_2\colon L_2 \rightarrow L$ one obtains $D_2$ and $H_2$. While $D_2$ is isomorphic to $K$, $\varphi_{R_2}\colon H_2 \rightarrow R$ is a monomorphism, but not an isomorphism. Rule $p_2$ thus agrees with $p$ in the \enquote{removal} phase, as they both delete the edge between the white and black vertices. However, the behavior in the \enquote{addition} phase is different, as $p_2$ leaves the white vertex disconnected as opposed to $p$, which creates an edge between the white and red vertices.
\end{example}

%% file: examples/input_transition_system.tex

\begin{example}[Input Transition System]
\label{ex:input_transition_system}

Here we present a small, loosely chemically inspired example showcasing the differences between the intended use of our input transition system and the standard use of a transition system to represent dynamics. The example consists of only three transitions, $\mc{T} = \{t_0, t_1, t_2\}$ as depicted in Figure~\ref{fig:input_transitions}. Figure~\ref{fig:transition_system}, in contrast, depicts the full transition system constructed from the initial graph $s_0$.

\begin{figure}[!ht]	
	\begin{subfigure}{0.45\textwidth}
		\centering

        \includegraphics[scale=1.2]{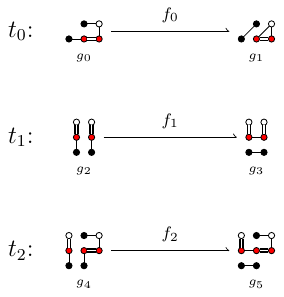}
		
		\caption{}
		\label{fig:input_transitions}
	\end{subfigure}
	\hspace{2em}
	\begin{subfigure}{0.45\textwidth}
		\centering

        \includegraphics[scale=1.2]{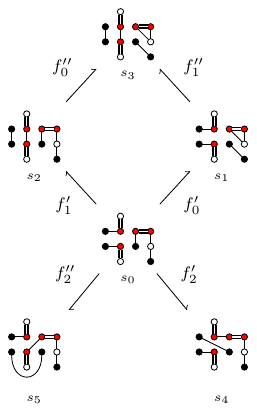}
		
		\caption{}
		\label{fig:transition_system}
	\end{subfigure}
	
	\caption{%
		An example transition system (\subref{fig:transition_system}) and the input transition system expressing equivalent behaviour (\subref{fig:input_transitions}). The relative position of the vertices is used to indicate the element maps. The element maps $f_0'$ and $f_0''$ are variations on the element map $f_0$ adapted to larger graphs. Same for $f_1$ and $f_2$.%
	}
	\label{fig:transitions}
\end{figure}

Observe that each of the three element maps appears twice in the full transition system (Figure~\ref{fig:transition_system}). This is caused by the \enquote{embedding} of a transition into larger graphs in order to represent full states. Indeed, the transformation encoded by the transition $t_0: g_0 \xrightharpoonup{f_0} g_1$ is trivially possible also for the graph $s_0 = g_0 \oplus g_2$, to produce $s_1 = g_1 \oplus g_2$, leaving the superfluous connected components (here $g_2$) untouched. A special case of such \enquote{embedding} is seen in the case of $f_2$, where the target graphs $s_4 \simeq s_5$ are isomorphic, however, the element maps $f_2' \not\equiv f_2''$ themselves are not. The input transition system (Figure~\ref{fig:input_transitions}) eliminates this duplication by only considering proper transitions.

So far we have only discussed the transitions themselves, that is the set $\mc{T}$, but not the set $\mc{U}$ of input graphs. Naturally, $\mc{U}$ has to contain at least the three connected components making up graphs $g_0, g_2$ and $g_4$. Assume that the transition system in Figure~\ref{fig:transition_system} is complete from the reachability perspective, i.e. no further transitions are possible from any of the graphs $s_0, \dots, s_5$. To reflect this, $\mc{U}$ must also contain the \enquote{product} components, or simply all connected components of the graphs $s_0, \dots, s_5$.

\end{example}

%% file: sections/methods.tex

\section{Methods}
\label{sec:methods}

In this section we introduce a formal method that takes an input transition system $S$ and produces a graph transformation model -- a set of rules $\mc{R}$, which can reproduce the behavior captured within $S$. To define this relationship formally, we introduce the concept of a \emph{generating} rule set. We begin with a single transition.

Let $t = G \xrightharpoonup{f} H\in \mc{T}(S)$ be a transition and $p$ a rule. We say that $p$ \emph{generates} the transition $t$, $p \models t$, if there exists a match $m\colon L(p) \rightarrow G$, such that the direct derivation $\delta = G \xrightarrow{m, p} H'$ has element map isomorphic to $f$, $f_{p(\delta)} \equiv f$. The set of all transitions generated by the rule $p$ sourced from the input graphs $\mc{U}(S)$ is then given as 
\[
p(\mc{U}(S)) = \{G\xrightharpoonup{f_{p(\delta)}} H\mid [G]\subseteq \mc{U}(S)\wedge \exists {m\colon L(p)\rightarrow G}, {\delta = G \xRightarrow{m,p}H} \}.
\]

Let now $\mc{R}$ be a set of rules. The definition of the set of all generated transitions naturally extends from single rules, $\mc{R}(\mc{U}(S)) = \bigcup_{p\in\mc{R}} p(\mc{U}(S))$. We say that $\mc{R}$ \emph{generates} the input transition system $S$, $\mc{R}\models S$, if $\mc{T}(S)\subseteq \mc{R}(\mc{U}(S))$, or equivalently, for each transition $t\in\mc{T}(S)$, there exists a rule $p\in\mc{R}$ such that $p$ generates $t$, $p\models t$. This definition of a generating rule set corresponds to the over-approximation case, where more transitions are produced by the rules than there are in the original input $S$. If a generating rule set $\mc{R}$ introduces no such \emph{spurious} transitions, $\mc{T}(S) = \mc{R}(\mc{U}(S))$, we say it is \emph{exact} for the input transition system $S$, $\mc{R} \vdash S$.

A single input transition system may have multiple different (exact) generating rule sets, as illustrated by Example~\ref{ex:generating_rule_sets}. Of special interest are the generating rule sets smallest with respect to the number of rules. Formally, a generating rule set $\mc{R}$ is considered \emph{minimal} (exact), if for any other (exact) generating rule set $\mc{R}'$ we have $|\mc{R}| \leq |\mc{R}'|$. Minimal rule sets not only correspond to the best possible data-compression ratio when we consider the problem as an instance of a graph transformation model compression, but are also crucial for determining the complexity of a graph transformation model.

\eject

Basing the minimality purely on the size of the rule set, each input transition system may have multiple different minimal (exact) generating rule sets. One could further refine the criterion by considering the size of the rules themselves, however, such refinements are beyond the scope of this publication.

\input{examples/generating_rule_sets}

\subsection{Maximum Rules}
\label{sec:maximum_rules}

One may note that the definition of a generating rule of a transition $t = G \xrightharpoonup{f} H$ is markedly similar to the definition of a subrule. Both rely on existence of a direct derivation which aligns with the original element map. This similarity is not coincidental. Consider a rule which corresponds precisely to the transition $t$, denoted by $p(t)$ and is only generating for the transition $t$, $p(t) \vdash t$. Intuitively, any rule which generates $t$ has to be a subrule of $p(t)$, making $p(t)$ the \emph{maximum} rule generating $t$.

\begin{figure}[!ht]
	\centering

    \includegraphics{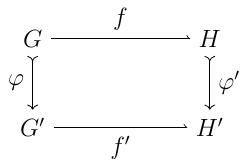}
 
	\caption{%
		A commuting diagram of two transitions, $G\xrightharpoonup{f} H$ and $G'\xrightharpoonup{f'}H'$ capturing the same transformation in different context.%
	}
	\label{fig:pathological_input}
\end{figure}

Depending on the input transition system $S$ and the transition $t\in \mc{T}(S)$, such a rule $p(t)$ might not exist. This happens in one particular case conditioned by existence of two graphs $G, G'$ composed of input graphs, $[G], [G'] \subseteq \mc{U}(S)$, and transitions $t= G\xrightharpoonup{f} H \in \mc{T}(S)$ and $t' =G'\xrightharpoonup{f'}H' \notin \mc{T}(S)$, such that $G \subseteq G'$ and such that the diagram in Figure~\ref{fig:pathological_input} commutes.

Since the DPO framework as introduced is unable to express negative application conditions, i.e. non-existence of an element, there are no means to disqualify $p(t)$ from applying to supergraphs of $G$. Furthermore, any subrules of $p(t)$ retain all of its applications, meaning any rule generating $t$ also generates $t'$, $p\models t\implies p\models t'$, and there exists no exact generating rule set for the input $S$. It should be noted that DPO itself has been extended to account for negative application conditions~\cite{EhrigEHP04}, the method can thus be adapted to settings where negative application conditions are allowed.

The following formal definition of the maximum rule $p(t)$ of a transition $t$ properly defines a rule for every transition, sacrificing instead the property of $t$ being the only generated transition. As a consequence, the method outputs a generating rule set which is not exact in case no such rule set exists, all spurious transitions being of the form of $t'$ from Figure~\ref{fig:pathological_input}.

\begin{definition}[Maximum Rule]
    Let $S$ be an input transition system and $t = G\xrightharpoonup{f} H \in \mc{T}(S)$ a transition. We now define the graph $K$ as the subgraph of $G$ on which $f$ is defined, $\dom(f)$, but with labels that disagree between $G$ and $H$ stripped.
    Formally, let $K$ be a graph and $g$ a surjective partial function $g\colon G \rightharpoonup K$ such that for all $x\in\dom(g) = \dom(f)$, $\lambda(g(x)) = \lambda(x)$ if $\lambda(x) = \lambda(f(x))$ and $\lambda(g(x)) = \epsilon$ otherwise.
    
    Then the maximum rule of the transition $t$ is a span $p(t) = (G \xleftarrow{l} K \xrightarrow{r} H)$ where $l = g^{-1}$ is a monomorphism by construction and $r = f\circ l$.
\end{definition}

Since $f_{p(t)} = r\circ l^{-1} = f \circ l \circ l^{-1} = f$ by definition, it becomes clear that $p(t)$ is generating for $t$, $p\models t$, by taking $\id_G$ as the match $m$. The set $\mc{R}(\mc{T}(S)) = \{p(t) \mid t\in \mc{T}(S)\}$ of all maximum rules is thus a generating rule set by definition, $\mc{R}(\mc{T}(S))\models S$. Example~\ref{ex:maximum_rules} illustrates such a maximum rule set.

\input{examples/maximum_rules}

\subsection{Candidate Rules}
\label{sec:candidate_rules}

While the maximum rules constitute a generating rule set, exact in case an exact generating rule set exists for the input transition system, it has by definition the same size as the transition set itself, $|\mc{T}(S)| =|\mc{R}(\mc{T}(S))|$, achieving no compression. Instead, $|\mc{R}(\mc{T}(S))|$ establishes the upper bound on the minimal generating rule set size, inclusion of any further rules being pointless for the purposes of recovering the original transitions.

To obtain a smaller generating rule set, we are thus interested in rules $p$ which are generating for at least two transitions $t_1, t_2\in\mc{T}(S)$, $p\models t_1,t_2$. As intuited above, any rule generating a transition, $p\models t$, is necessarily a subrule of the maximum rule $p\subseteq p(t)$. The subrules of the maximum rules are thus \emph{candidate} rules for the minimal generating rule set. Or alternatively, the set of all candidate rules is the downward closure of $\mc{R}(\mc{T}(S))$ with respect to the subrule relation, ${\downarrow}\mc{R}(\mc{T}(S))$. The nature of the candidate rule space is explored in Example~\ref{ex:candidate_rules}.

\input{examples/candidate_rules}

The problem of constructing a minimal generating rule set from among the candidate rules reduces to an instance of the set cover problem~\cite{Vazirani03_ch2}. In particular, the transition set $\mc{T}(S)$ of the input transition system is the universe to be covered and each candidate rule $p$ is equipped with a coverage $c\colon {\downarrow}\mc{R}(\mc{T}(S)) \rightarrow 2^{\mc{T}(S)}$ defined as $c\colon p \mapsto \left\{t\in\mc{T}(S)\mid p\models t\right\}$. Finding a minimal rule set $\mc{R}\subseteq {\downarrow}\mc{R}(\mc{T}(S))$ is then exactly searching for the smallest subset $C \subseteq {\{c(p)\mid p\in{\downarrow}\mc{R}(\mc{T}(S))\}}$ such that $\bigcup C = \mc{T}(S)$.

The set cover problem is well known to be NP-hard~\cite{Karp72}, making the number of candidate rules (sets) a concern. A property which does not mix well with the combinatorial nature of the candidate rule space. The coverage function $c$ is discrete and defines an equivalence relation ${\equiv_c} \subseteq {\downarrow}\mc{R}(\mc{T}(S)) \times {\downarrow}\mc{R}(\mc{T}(S))$ on the candidate rules, such that two rules $p_1 \mathbin{\equiv_c} p_2$ are equivalent if and only if they have equal coverage, $c(p_1) = c(p_2)$. Rules belonging to the same equivalence class are indistinguishable from the perspective of the set cover problem, which can be executed on the equivalence classes themselves, rather than the individual rules, greatly reducing the number of candidate sets to consider.

Moreover, a subrule $p_1 \subseteq p_2$ is by definition guaranteed to retain all possible applications of $p_2$. In particular, $c(p_1) \supseteq c(p_2)$. It follows that for any generating rule set $\mc{R}$ and for any $p'\subseteq p$ such that $p\in\mc{R}$, the rule set $(\mc{R} \setminus \{p\}) \cup \{p'\}$ is also generating. A further optimisation is therefore possible by only considering the equivalence classes of the minimal candidate rules.

For input transition systems which have an exact generating rule set as discussed above, the minimal exact generating rule set can be obtained via the same procedure, the difference being the candidate rules to consider. In particular, only rules which introduce no spurious transitions, $\{p\in {\downarrow}\mc{R}(\mc{T}(S)) \mid p(\mc{U}(S)) \subseteq \mc{T}(S)\}$, should be considered for an exact generating rule set. Such a minimal exact generating rule set $\mc{R}$ also provides a measure of complexity of the input transition system, $|\mc{R}|$ being the smallest number of rules one needs to recreate exactly the transitions in $\mc{T}(S)$. The complexity viewpoint is discussed in greater detail in Section~\ref{sec:complexity}.

Finally, the minimality of the generating rule set, or the data-compression ratio, is not the only parameter of interest in the lossy compression scenario. In particular, the minimal generating rule set $\mc{R}$ might in some cases have very high distortion rate, given by the number of spurious transitions, $|\mc{R}(\mc{U}(S))\setminus\mc{T}(S)|$. A generalization of the set cover problem is needed to optimize for both parameters, the data-compression and distortion rates.

Just as the coverage being greatest for minimal subrules, so is the amount of generated spurious transitions. The equivalence classes of minimal candidate rules are therefore not sufficient when considering distortion rate. Moreover, rules belonging the same equivalence class by coverage might generate different spurious transitions. Consider now $p_1 \subseteq p_2\in[p_1]_{\equiv_c}$. Once again we know that any spurious transitions generated by $p_2$ will also be generated by $p_1$. It is therefore sufficient to consider the maximal rules in each equivalence class $\mc{R}_c=\{{p \in {\downarrow}\mc{R}(\mc{T}(S))} \mid \forall p'\in[p]_{\equiv_c}, p\not\subset p'\}$.

A well established solution of the set cover problem uses integer linear programming (ILP)~\cite{Vazirani03_ch13}. In the following we provide a generalization of the ILP formulation of the set cover problem whose objective function~(\ref{eq:objective_function}) penalizes both number of rules and number of spurious transitions. This is achieved by considering three types of variables to represent the candidate rules, input transitions and spurious transitions.

\begin{equation}
    O = \sum_{p \in \mc{R}_c}x_{p} + \sum_{t \in T}z_{t}
    \label{eq:objective_function}
\end{equation}
where $x_{p}$ for each $p\in \mc{R}_c$ is a variable representing the candidate rule, 
\[
T = ({\downarrow}\mc{R}(\mc{T}(S)))(\mc{U}(S)) \setminus \mc{T}(S)
\]
 is the set of all possible spurious transitions and $z_t$ for each $t\in T$ is a variable representing the spurious transition, accompanied with a constraint $z_t \geq x_p$ for all rules $p\in\mc{R}_c$ generating the transition $t$, $p\models t$.

Additionally, to ensure that the produced rule set is indeed generating, a variable $y_t$ for each $t\in\mc{T}(S)$ is included, representing the input transitions. Each such variable is accompanied with the following constraints, $y_t \geq 1$ and $y_t = \sum_{\mc{R}_c\ni p\models t}x_{p}$.

The desired relationship between the data-compression and distortion rates can then be regulated by including scaling factors to determine how many rules have to be eliminated by a spurious transition for it to be included, or conversely, how many spurious transitions are permissible for the elimination of one rule. The precise control of the relationship between the two parameters is instrumental for suggesting completion of the input transition system by means of lossy compression.

\subsection{Kolmogorov Complexity of Graph Transformation Models}
\label{sec:complexity}

As aforementioned, the sizes of minimal exact generating rule sets define a complexity metric on the input transition systems and by extension, the graph transformation models whose behavior they represent. Formally, we write $K(S) = \min_{\mc{R}\vdash\mc{T}(S)}(|\mc{R}|)$ to denote the complexity of the input transition system $S$.

The complexity $K(S)$ measures the number of distinct transformations, including their execution conditions, allowed by the graph transformation model. The measure $K(S)$ thus captures the amount of elementary behaviors the model exhibits, or in other words, the diversity of the model dynamics. Such a complexity measure is useful for study and evaluation of empirically inferred input transition systems. In the case of chemical reaction networks, for instance, the complexity allows for formal reasoning about the underlying model of chemistry, opening up possibilities for comparison and further analysis of various industrial and biological processes.

The complexity measure is also of interest from the computational perspective. Akin to a formal grammar in formal language theory, a graph transformation model can be treated as a solution of a computable problem. $K(S)$ then measures the complexity of said solution (or program). Rather than measuring the resources required for computation, such as time or space, $K(S)$ measures the smallest encoding of the solution, or program, itself, closely resembling Kolmogorov complexity~\cite{Kolmogorov98}, which measures the minimum size of a program solving a given problem. Indeed, the minimal exact generating rule set based complexity is an abstraction of the Kolmogorov complexity adapted to graph transformation models. Abstraction on the account of only considering the number of rules, not the size of the rules themselves.

A reader learned in complexity theory might correctly remark that Kolmogorov complexity of a computable problem is undecidable~\cite{ChaitinAC95}. Indeed, $S$ is only one graph transformation solution of a given problem and $K(S)$ only expresses the complexity of that single solution, not the minimum across the entire solution (program) space. This disparity is illustrated in Example~\ref{ex:edge_contraction} using edge contraction to compute a spanning tree.

\input{examples/edge_contraction}

%% file: examples/generating_rule_sets.tex

\begin{example}[Generating Rule Sets]
\label{ex:generating_rule_sets}

Consider the input transition system $S$ from Example~\ref{ex:input_transition_system}. In spite of only having three transitions, there exist multiple rule sets (exact) generating $S$. We showcase some of them in Figure~\ref{fig:generating_rule_sets}.

\begin{figure}[!ht]	
	\begin{subfigure}{0.31\textwidth}
		\centering

        \includegraphics[scale=1.2]{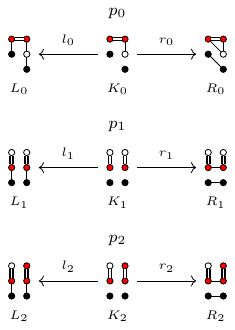}
		
		\caption{}
		\label{fig:generating_rule_set}
	\end{subfigure}
	\hspace{1em}
	\begin{subfigure}{0.31\textwidth}
		\centering

        \includegraphics[scale=1.2]{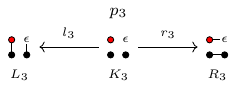}
		
		\caption{}
		\label{fig:minimal_not_exact}
	\end{subfigure}
	\hspace{1em}
	\begin{subfigure}{0.31\textwidth}
		\centering

        \includegraphics[scale=1.2]{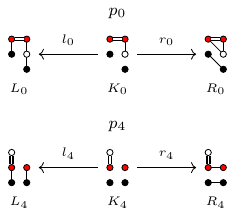}
		
		\caption{}
		\label{fig:minimal_semantics}
	\end{subfigure}
	
	\caption{%
        Example of three generating rule sets for the input transition system from Example~\ref{ex:input_transition_system}. Unlike the rule set in (\subref{fig:minimal_not_exact}), both of the rule sets in (\subref{fig:generating_rule_set}) and (\subref{fig:minimal_semantics}) are exact. (\subref{fig:minimal_semantics}) is additionally a smallest exact rule set generating the input transition system.%
	}
	\label{fig:generating_rule_sets}
\end{figure}

Let us first consider the rule set $\{p_0, p_1, p_2\}$ (Figure~\ref{fig:generating_rule_set}). This rule set uses a unique generating rule for each of the transitions, $p_0$ for $t_0$, $p_1$ for $t_1$ and $p_2$ for $t_2$, and is exact, allowing no spurious behavior. Upon closer inspection, however, we observe each of the rules performs the same underlying transformation, differing only in the unchanging (invariant) part. The rules $p_0, p_1, p_2$ having a common subrule, indicates that a smaller generating rule set might exist.

Indeed, the singleton set $\{p_3\}$ (Figure~\ref{fig:minimal_not_exact}) is a minimal generating rule set, however, it generates spurious transitions. E.g., $p_3$ can be applied to $G = g_0 \oplus g_0$ to create a transition not present in $\mc{T}(S)$.

Ultimately, at least two rules are required to obtain an exact generating rule set, $\{p_0, p_4\}$, making it the minimal exact (Figure~\ref{fig:minimal_semantics}).

\end{example}

%% file: examples/maximum_rules.tex

\eject

\begin{example}[Maximum Rules]
\label{ex:maximum_rules}

Consider the input transition system $S$ as depicted in Figure~\ref{fig:open_lts} with the input graph set $\mc{U}(S) = \{g_0, \dots, g_5\}$ containing all the graphs. Since all the graphs are connected, the input transition system $S$ coincides with the usual notion of a transition system and is depicted accordingly.

\begin{figure}[!ht]
    \begin{subfigure}{1\textwidth}
        \centering

        \includegraphics[scale=1.2]{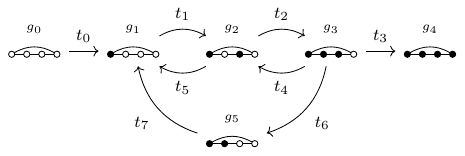}
   
        \caption{}
        \label{fig:open_lts}
    \end{subfigure}
    \begin{subfigure}{1\textwidth}
    	\centering

        \includegraphics[scale=1.2]{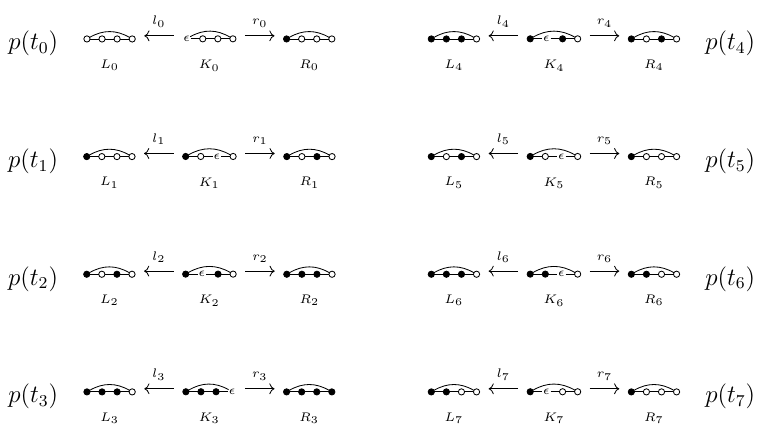}
     
        \caption{}
        \label{fig:maximum_rule_set}
    \end{subfigure}
	
	\caption{%
        An input transition system $S$ (\subref{fig:open_lts}) and the maximum rule set of $\mc{R}(\mc{T}(S))$ (\subref{fig:maximum_rule_set}). The element maps are implied by relative vertex positions.%
	}
	\label{fig:maximum_rules}
\end{figure}

The maximum rule $p(t)$ for each of the eight transitions $t\in \mc{T}(S)$ is given in Figure~\ref{fig:maximum_rule_set}. The transitions in this example are simple, each only changing the color (label) of a single vertex. This translates into empty labels in the invariant graphs $K_0, \dots, K_7$.

\end{example}

%% file: examples/candidate_rules.tex

\begin{example}[Candidate Rules]
\label{ex:candidate_rules}

Consider the input transition system $S$ from Example~\ref{ex:maximum_rules}. Examining the maximum rule set, one can identify two groups of transitions, ones that recolor a vertex from white to black $t_0, t_1, t_2$ and $t_3$, and the ones that perform the opposite operation, recoloring from black to white $t_4, t_5, t_6$ and $t_7$. The transformations represented by these two groups of transitions are fundamentally different, the respective maximum generating rules sharing no common subrules and splitting the candidate rule space into two disjoint subsets, the subrules of $p(t_0), \dots, p(t_3)$ and subrules of $p(t_4), \dots, p(t_7)$. For the sake of simplicity, we limit ourselves to ${\downarrow}\{p(t_4), \dots, p(t_7)\}$ within this example. We further only consider candidate rules with all vertices labeled, outside of the invariant graph $K$, as the identical topology of all the input graphs makes unlabeled vertices effectively identical to no vertices.

The candidate rule space ${\downarrow}\{p(t_4), \dots, p(t_7)\}$ is given in Figure~\ref{fig:candidate_rules} arranged into a Hasse diagram of the subrule relation. To save space, we draw each rule as a single graph constructed as the union of the left and right side. The vertices changing label are colored half-and-half in their old and new colors. Each rule $p$ is additionally labeled with its coverage $c(p)$, as defined further in the text.

\begin{figure}[!ht]
	\centering

    \includegraphics[scale=1.2]{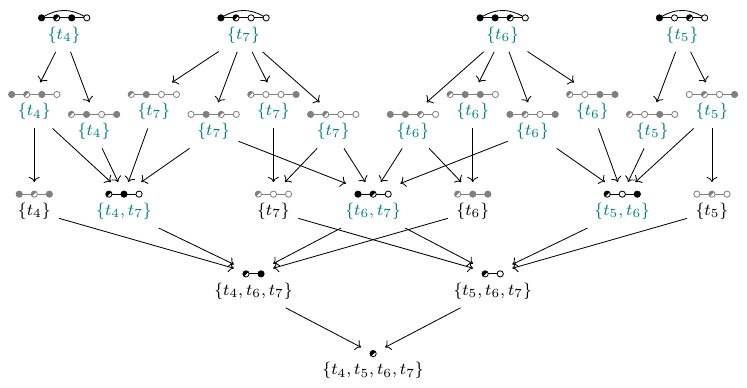}

	\caption{The candidate rules for a vertex turning white in the input transition system from Example~\ref{ex:maximum_rules}. The candidate rules are arranged into a Hasse diagram with the subrule relation. Rules which are not maximal common subrules of a subset of the maximal rules are grayed out. The coverage of each rule is used as a label, highlighted in teal color if they introduce no spurious transitions.}
	\label{fig:candidate_rules}
\end{figure}

The four rules at the top of Figure~\ref{fig:candidate_rules} are the maximum rules for each of the transitions. In this example, all the rules share a single minimal subrule, at the bottom of the diagram. This minimal rule thus constitutes the solution for a minimal generating rule set. The minimal exact generating rule set instead consists of the rules covering $\{t_4, t_7\}$ and $\{t_5, t_6\}$ in the middle row, as they produce no spurious transitions. Both of these solutions can be obtained by considering only maximal common subrules of the four maximum generating rules. This is indicated by the other rules being grayed out.

Notice that while the example resembles cellular automata reduced to one dimension, the behavior cannot be expressed in the standard cellular automata fashion, where the value of each agent depends only on its immediate neighbors. Instead, the exact minimal generating rule set consists of two rules which consider two closest neighbors in a single direction.

\end{example}

%% file: examples/edge_contraction.tex

\begin{example}[Edge Contraction]
\label{ex:edge_contraction}

In this example, we consider the simple problem of constructing a spanning tree of simple graphs without edge weights (unlabeled), $|{}^{\bullet}e| = |e^{\bullet}| = 1$ for all edges $e \in E(G)$ and $\Sigma=\emptyset$. A spanning tree of such a graph can be encoded by a repeated application of the edge contraction operation as described in~\cite{Rosen11}. Intuitively, contraction of an edge $e$ consists in deleting the edge alongside both of its endpoints $u$ ($\{u\} = {}^{\bullet}e$) and $v$ ($\{v\} = e^{\bullet}$) and replacing it by a single vertex $w$ which inherits all the edges of both $u$ and $v$ except $e$ itself. Contracting a graph into a single vertex defines a spanning tree of the graph by means of the edges that were contracted.

\begin{figure}[!ht]
    \begin{subfigure}{0.5\textwidth}
        \centering

        \includegraphics[scale=1.2]{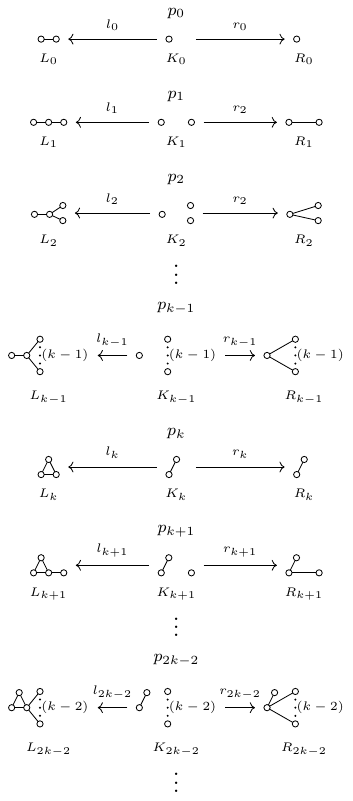}
        
        \caption{}
        \label{fig:exhaustive_contraction}
    \end{subfigure}
    \hspace{2em}
    \begin{subfigure}{0.5\textwidth}
        \centering
        
        \includegraphics[scale=1.2]{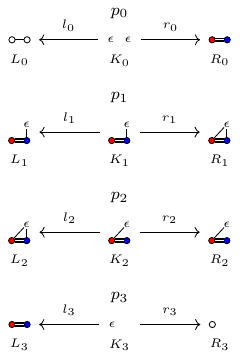}
        
        \caption{}
        \label{fig:stepwise_contraction}
    \end{subfigure}
	
	\caption{%
        The minimal exact generating rule sets of the two input transition systems $S_1$ and $S_2$ modeling spanning tree construction by edge contraction. (\subref{fig:exhaustive_contraction}) Illustration of the rule set for $S_1$ which does one edge contraction per transition. The size of the rule set depends on the input graphs in $\mc{U}(S_1)$. (\subref{fig:stepwise_contraction}) The entire rule set for $S_2$, consisting of only four rules thanks to use of \enquote{non-terminal} labels.%
	}
	\label{fig:edge_contraction}
\end{figure}

Consider two different ways of encoding edge contraction as a graph transformation model, or more precisely an input transition system. Both input transition systems $S_1$ and $S_2$ operate under the same principle, only one vertex is deleted, the other endpoint assuming the role of the new vertex $w$. The edge contraction itself is thus performed by reconnecting all edges of the deleted endpoint to the remaining one. The difference between the two systems being that in $S_1$ all edges are reconnected at the same time, one transition corresponding to exactly one edge contraction. $S_2$ on the other hand, reconnects each edge individually, first by duplicating it for the remaining endpoint (if necessary) and then deleting it. Such an approach also requires use of vertex and edge labels to keep track of the edge being contracted, akin to the use of non-terminal symbols in formal grammars.

Although $S_1$ and $S_2$ solve the same underlying problem, their respective minimal exact generating rule sets might be of vastly different size, as illustrated in Figure~\ref{fig:edge_contraction}. Since each edge contraction is performed within a single transition in $S_1$, a different rule is necessary depending on the degree of the deleted vertex. Additionally, a different rule is required for all possible numbers of neighbors shared by both endpoints of the contracted edge to avoid creating a multiple edges. Assume $k$ is an upper bound on vertex degrees, $k = \max_{G\in \mc{U}(S_1)}(|V(G)|-1)$, the minimal exact generating rule set is then in the worst case (all possible combinations of neighbours represented within the input graphs) of size $\frac{k}{2}(k + 1)$. Thus in particular, $K(S_1)$ depends on the size of the input (largest vertex degree). Figure~\ref{fig:exhaustive_contraction} illustrates the shape of the necessary rules.

The minimal exact generating rule set of $S_2$ (Figure~\ref{fig:stepwise_contraction}), on the other hand, is of constant size, $K(S_2) \leq 4$, irrespective of the size of the input graphs. The Kolmogorov complexity of spanning tree construction by edge contraction is therefore at most four and surely constant. An observation impossible to conclude from $S_1$ alone, illustrating how heavily the introduced compression depends on the exact modeling of the underlying problem -- a solution.

\end{example}

%% file: sections/experiments.tex

\section{Experiments}
\label{sec:experiments}

In this section, we introduce several, mostly artificial, application examples significantly more complex than the input transition system used in Example~\ref{ex:maximum_rules}. All experiments were performed using a prototype implementation (\url{https://github.com/JuriKolcak/rule_inference}) tailored for use in concert with the graph transformation framework MØD. MØD is a graph transformation framework specialized for chemical reaction networks and implements DPO in accordingly constrained fashion. In particular, only simple graphs are considered and the right morphisms $r$ and matches $m$ of all rules are required to be injective (monomorphisms)~\cite{AndersenFMS16}.
It should be noted that the current implementation is purely a proof of concept to illustrate the application possibilities of the method and is not the focus of the publication or at all optimised for performance. The implementation of many sub-problems, such as finding of the maximal common subrule, is strictly naive. For this reason, we do not consider computational performance evaluation as part of the experiments.

\subsection{Regular Language}
\label{sec:regular_grammar}

In this experiment, we reproduce a simple formal grammar encoding a regular language within the graph transformation framework. The grammar chosen has three non-terminal symbols ($S, A, T$), two terminal symbols ($0, 1$) and seven rules listed below. The encoded formal language is equivalently specified by the regular expression $(0|10)^*1^*$. The grammar has been chosen to showcase the importance of proper encoding within the graph transformation framework.

\begin{center}
    \begin{tabular}{llllll}
        $S \rightarrow AT$ & (1)\qquad~ & $A \rightarrow \epsilon$ & (2)\qquad~ & $T \rightarrow \epsilon$ & (6) \\
         & & $A \rightarrow AA$ & (3) & $T \rightarrow T1$ & (7) \\
         & & $A \rightarrow 0$ & (4) & & \\
         & & $A \rightarrow 10$ & (5) & &
    \end{tabular}
\end{center}

The simplest encoding is representing the string by simple undirected path graphs with the vertex labels $\Sigma_0=\{S, A, T, 0, 1\}$ representing the characters of the string. We use this encoding for the input transition system $S_0$. As the regular language generated by our grammar is infinite, we only include graphs representing strings of up to length five in $\mc{U}(S_0)$ and all transitions allowed for any such word in $\mc{T}(S_0)$. The transitions are modelled by deletion of the vertex representing a non-terminal symbol and placement of vertices encoding the right hand side of the rule in its stead, so that the path graph structure is preserved.

Owning to the simplicity of the encoding used to construct $S_0$, one does not recover the seven rules. Instead, no exact generating rule set exists for $S_0$. This is due to an instance of the condition depicted by the diagram in Figure~\ref{fig:pathological_input}. In particular, the rule rewriting the non-terminal $A$ to a couple of terminals $10$ (5) is order-dependent. In the undirected graph setting, however, the maximal rule $p(t)$ of the transition $t = AA0 \xrightharpoonup{\{0\mapsto 0, 2\mapsto 3\}} A100 \in \mc{T}(S_0)$ can be applied to the graph representing the string $10AA$ in \enquote{reverse order} to produce a transition $t' = 10AA \xrightharpoonup{\{0\mapsto 0, 1\mapsto 1, 3\mapsto 4\}} 1001A \notin \mc{T}(S_0)$.

Let thus $S_1$ be another input transition system, encoding the grammar by directed paths instead. Since only vertices are labelled, the label set $\Sigma_1=\Sigma_0$ is the same as for $S_0$. $\mc{U}(S_1)$ and $\mc{T}(S_1)$ again contain all the relevant graphs and transitions representing strings up to length five. $S_1$ does indeed have an exact generating rule set, however, the minimal exact generating rule set is of size $25$, almost four times as many as the original seven rules. This is no coincidence. The transitions require that the non-terminal vertices be deleted and replaced by new ones, encoded by the absence of the relevant elements in the element maps. Each of the regular grammar rules (save for the rule rewriting the initial non-terminal $S$ (1)) thus have to be represented by four graph transformation rules depending on their adjacent edges; none, if the string is of length one; one on the right, if the rule applies to the start of the string; one on the left, if the rule applies to the end of the string; or both, if the rule applies in the middle of the string.

To resolve the above issue, we propose a third, and final, encoding within the input transition system $S_2$. $S_2$ is identical to $S_1$ with the exception that each graph has two extra vertices labeled as $\triangleright$ and $\triangleleft$, marking the start and the end of the string, respectively. The label set $\Sigma_2 = \Sigma_1 \cup \{\triangleleft, \triangleright\}$ is thus larger to accommodate the new labels. These vertices allow all rules to expect edges on both sides, irrespective of where in the string are they applied. $\mc{U}(S)$ and $\mc{T}(S)$ still represent all the strings up to length five, resulting in path graphs up to length seven including the \enquote{start} and \enquote{end} symbols, and the relevant transitions. The minimal exact generating rule set of $S_2$ is indeed of size seven, recovering precisely the original formal grammar rules.

It should be noted that the string length cut-off of five has been chosen arbitrarily as \enquote{sufficiently long}. In fact, the seven rules can be retrieved already if only strings of length one, representing the non-terminal symbols, are considered. However, depending on the encoding, the obtained rules would not be applicable for strings of arbitrary length and one would fail to capture the interesting side-effects of the graph representation induced by neighbouring vertices, such as the quadruple blow-up of the number of rules in the $S_1$ case.

\subsection{Tic-Tac-Toe}
\label{sec:tictactoe}

In this example, we look at the classical game of Tic-Tac-Toe, played on a three-by-three playing board. In particular, we infer the rules which make up a winning, or rather a non-losing, strategy of either player.

Formally, the game of Tic-Tac-Toe is represented by simple undirected graphs with the label set $\Sigma=\{\_, O, X\}$ only containing three labels, $\_$ representing an empty tile, $O$ a tile claimed by player one and $X$ a tile claimed by player two. The graphs represent game states, making each graph of fixed size, with nine vertices for the tiles and twelve edges between neighbors. The transition system representing the game dynamics consists of the players alternating in placing one copy of their symbol into an empty tile, beginning with player one. The game ends when a player connects three symbols vertically, horizontally or diagonally, thus winning, or if victory is no longer possible for either player (tie). Additionally, we demand all vertices to be labelled in this experiment, due to the combinatorial complexity, with the exception of the invariant graph $K$ of the rules.

\begin{table}[!t]
    \centering
    \begin{tabular}{|l||r|r|r|r|}
        \hline
         & $|\mc{U}|$ & $|\mc{T}|$ & cand. rules & $|\mc{R}|$ \\
        \hline
        Player one ($S_O$) & \num{311} & \num{684} & \num{11419} & \num{157} \\
        Player two ($S_X$) & \num{222} & \num{339} & \num{3737} & \num{89} \\
        \hline
    \end{tabular}
    \caption{The number of input graphs and transitions, the computed candidate rules (only maximal common subrules are considered) and the size of the minimal exact generating rule set for both input transition systems $S_O$ and $S_X$.}
    \label{tab:player_numbers}
\end{table}

\begin{figure}[!t]
    \centering
    
    \includegraphics{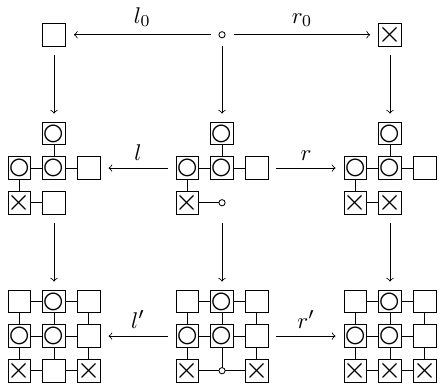}
    
    \caption{The middle row is an example rule from the minimal exact generating rule set of player 2 ($S_X$). The unique minimal rule is depicted at the top, while the maximum rule of one of the \num{13} generated transitions is shown at the bottom.}
    \label{fig:tictactoe_rule}
\end{figure}

In this example, we look at player-specific models, considering two input transition system, $S_O$ for player one and $S_X$ for player two. Only graphs representing game states in which player one makes their move are included in $\mc{U}(S_O)$ and similarly, only states in which player two plays are included in $\mc{U}(S_X)$. Finally, since we are interested in non-losing strategies, only transitions which deny victory to the opposing player, i.e. do not lead to a state in which the opponent has a winning strategy, are considered in either $\mc{T}(S_O)$ and $\mc{T}(S_X)$. Similarly only states reachable by such transitions are in $\mc{U}(S_O)$ and $\mc{U}(S_X)$, i.e. a state in which player one plays, but which is winning for player two, is not considered in $\mc{U}(S_O)$ and vice-versa.

Due to the asymmetric nature of the game, the starting player, player one, has more possibilities for victory. This is reflected in the number of transitions in $\mc{T}(S_O)$ being significantly higher than in $\mc{T}(S_X)$, as well as $\mc{U}(S_O)$ containing more graphs (game states) than $\mc{U}(S_X)$. The exact counts can be found in Table~\ref{tab:player_numbers}.

The game of Tic-Tac-Toe provides a very unique landscape for our method since all transitions share the same minimal rule. The minimal generating rule set for either player is thus of size one. Such rule set, however, corresponds to all the legal moves either of the players can make, rather than the coveted non-losing strategy, which requires an exact generating rule set. As indicated in Table~\ref{tab:player_numbers}, encoding the non-losing strategy of either player as graph transformation rules is relatively complex, resulting in relatively large rule sets. This is reflected by the shape of the rules themselves, each of which specifies a significant portion of the playing board (example in Figure~\ref{fig:tictactoe_rule}). Nonetheless, the rule sets are still non-trivially smaller than the explicit list of all transitions belonging to either player's strategy.

\subsection{Organic Chemistry -- Formose Reaction}

Last but not least, we consider an example from chemistry. In particular, we look at the well-studied mechanism of monosaccharide synthesis from formaldehyde, commonly known as the formose reaction~\cite{Butlerov61, Breslow59, Cleaves11, AppayeeB14, DelidovichSTP14}. The formose reaction is a suitable and concise example, but above all highly relevant in chemistry, especially in the study of auto-catalysis~\cite{Socha80} and origin of life studies~\cite{RicardoCOB04, BennerKKR10, PallmannSHLHT18}.

\begin{figure}[!ht]
	\centering

    \includegraphics{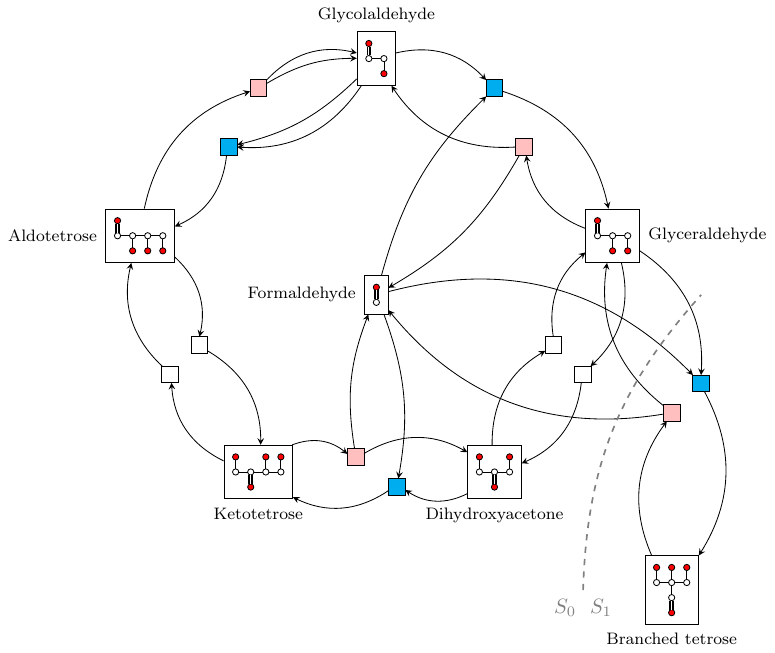}
    
    \caption{%
        The schema of the formose reaction doubling as the input transition systems $S_0$ and $S_1$. The transitions are depicted in a Petri net-style, each connected component of the source and target graphs linked with a separate arc. The transitions are colored by the reaction type represented. Blue for aldol reactions, pink for the reverse retroaldol reactions and white for aldose-ketose isomerization reactions. The part separated by the gray dashed line in the bottom right, which exemplifies the production of branched monosaccharides, only appears in the input transition system $S_1$.%
    }
    \label{fig:formose_reaction}
\end{figure}

Molecules are easily represented as simple graphs, with the vertices standing in for atoms and edges for bonds between them, both typed by the labeling function. In the context of the formose reaction, carbon, oxygen and hydrogen atoms, as well as single and double bonds are sufficient, giving us the following label set $\Sigma=\{C, O, H, {-}, {=}\}$. The chemical reactions then become the transitions of the input transition system. For the formose reaction itself, we consider the main auto-catalytic cycle~\cite{Breslow59, Cleaves11, DelidovichSTP14}, consisting of 5 reversible steps (10 transitions) for the input transition system $S_0$, as depicted in Figure~\ref{fig:formose_reaction}.

Although explicit enolization step has been often considered for the formose reaction as a precursor for both the aldol reaction and isomerisation steps~\cite{BennerKKR10, AndersenFMS14}, the isomerisation step has been demonstrated to proceed via a hydride shift mechanism instead of enolization~\cite{AppayeeB14}. We therefore omit explicit enolization transitions, encoding the enolization step within the aldol and retroaldol reactions implicitly.

The minimal exact generating rule set for the input transition system $S_0$ contains six rules. The aldose-ketose isomerization reactions are represented by two rules while a single rule is sufficient for the retroaldol reactions. The final three rules generate the three aldol reactions, meaning the respective maximum rules of each of the transitions have been used.

It is well understood that a wide variety of monosccharides of various sizes are synthesized during the formose reaction, which do not appear in the core auto-catalytic cycle (Figure~\ref{fig:formose_reaction}). We thus repeat the experiment with a scaling factor $\rho$ used to multiply the size of the rule set in the objective function of the ILP, thus encouraging inclusion of spurious transitions in order to constrain the size of the rule set.

Already at $\rho = 2$ (adding rules is twice as expensive as allowing spurious transitions), we retrieve a rule set of size five at the cost of one spurious transition. At $\rho = 3$ (rules triple as expensive) the resulting rule set is of size four at the price of another two spurious transitions for a total of three. Finally, at $\rho >= 9$ (rules at least nine times as expensive) we retrieve only three rules, one for each type of reaction, with a total of eleven spurious transitions.

The spurious transitions are generated by two of the retrieved rules. Two generated by the rule performing aldose-ketose isomerization and nine generated by the aldol reaction rule. The spurious transitions of the aldose-ketose isomerization are \enquote{self-isomerizations}, that is glycolaldehyde or ketotetrose undergoing aldose-ketose isomerisation into themselves, that is the result graphs of the rule application are isomorphic to the input graphs. This type of isomerization might be important for tracing individual atoms across reactions or for stereochemistry, however, is superfluous, albeit not incorrect, at the level of abstraction we employ in this example.

\eject

The remaining nine spurious transitions, generated by the rule representing the aldol reaction, correspond to production of larger monosaccharides, incl. aldo-{}, 2-keto-{} and 3-keto-{} pentoses and hexoses, two heptoses and even an octose. All such monosaccharides are among the expected products of the formose reaction. Indeed, upon applying the rules repeatedly, we retrieve also the backwards retroaldol reactions and isomerisations of the produced monosaccharides as well as further productions, all in line with the expected chemistry of the formose reaction~\cite{Breslow59, Cleaves11, DelidovichSTP14, PallmannSHLHT18}. The inferred rules thus constitute a successful model completion starting just from the core auto-catalytic cycle.

It should be noted that the rules we obtain are not minimal, even excluding empty labels ($\epsilon$). In fact, the completion of the model we obtain is still too restrictive. It is, for instance, unable to produce branched monosaccharides (molecules whose carbon atoms aren't arranged in a path graph), known to also appear among the numerous products of the formose reaction. To attempt recovery of the production of branched monosaccharides, we consider a second input transition system $S_1$, equal to $S_0$ enriched by a branched tetrose and the aldol and retroaldol reactions producing it and breaking it up, respectively (Figure~\ref{fig:formose_reaction}).

The minimal exact generating rule set of $S_1$ contains seven rules. The only difference from $S_0$ being a new rule generating the additional aldol reaction, producing the branched tetrose. The similarity to $S_0$ persists at $\rho = 2$, where a rule set of size six is retrieved at the cost of one spurious transition. At $\rho = 3$ we already get a rule set of size only four, producing a total of five spurious transitions, including production of some branched pentoses. Finally, we retrieve a generating rule set of size three at $\rho >= 24$, generating a total of $28$ spurious transitions. All the spurious transitions are indeed productions of various monosaccharides, this time including the branched variations, showcasing the combinatorial nature of the formose reaction.

\begin{figure}[!ht]
    \centering

    \includegraphics{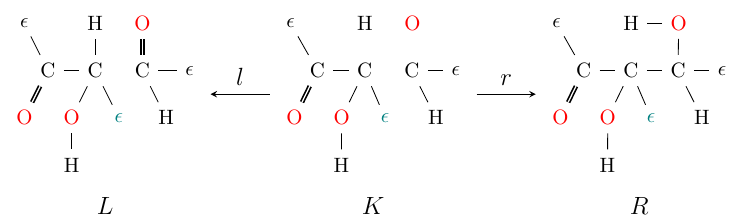}
    
    \caption{%
        The rule encoding the aldol reaction in the minimal generating rule set of $S_1$. The analogous rule in the minimal generating rule set of $S_0$ explicitly marks the bottom unlabeled vertex (highlighted in teal) as a hydrogen (\ch{H}), thus forcing the middle carbon to be \enquote{at the end} of a molecule and preventing creation of branched monosaccharides. The same difference is present in the rules generating the retroaldol reactions.%
    }
    \label{fig:aldol_reaction}
\end{figure}

Interestingly, the rules representing the aldose and retroaldose reactions in the minimal generating rule sets of $S_0$ and $S_1$ only differ in a single vertex. Or more precisely its label, as illustrated in Figure~\ref{fig:aldol_reaction}. The rule also shows that a considerably large part of the graphs (molecules) outside of the directly modified edges is preserved (the invariant graph $K$). Such \enquote{context} represents the structure shared by all four aldol reactions in $S_1$.

%% file: sections/discussion.tex

\section{Discussion}
\label{sec:discussion}

Design of graph transformation models -- or graph transformation rules -- is dominated by manual approaches. We present a formal method which can be employed to automate, at least partially, the rule design process. The method is versatile and can be applied at different stages of the modeling process to perform or aid with various tasks, including construction of an initial design, verification, completion, evaluation or optimization of the graph transformation models.

The method is presented on graphs, but the only required formalisms, the notion of a subrule and the ability to express proper transitions (coproducts), allows it be extended to any DPO application and, in particular, extensive adhesive categories in general, modulo the representation of the element map. The method is therefore highly adaptable to the exact application setting, opening up possibilities for various optimizations. The adaptability is further enhanced by the possibility to freely augment the objective function of the ILP instance making up the final step of the procedure.

The generality of the method is underlined by demonstration of the rule inference method on multiple highly varied models. The experiments serve not only to highlight the method itself, both for inference of exact generating rule sets and model completion, but also to exemplify the outstanding challenges. In particular, the method is reliant on a precise specification of the input transition system, which is not always the case when working with empirical data.

Our contribution thus lies mainly in the formalization of the framework which allows us to explore graph transformation models from the perspective of existing, empirical or desired, behavior. The precise implementation, which is merely one of the numerous possibilities, takes on a background role. Even as a formal framework, we only dissect part of the problem -- the over-approximation -- omitting the under-approximation case, that is allowing some of the transitions of the input transition system to not be generated, in favor of more careful treatment.

Our formalism attempts to shed light on a new approach to the formal analysis of graph transformation, as well as other rule-based and generative models as a whole. We believe the new perspective in combination with the flexibility of the method makes for a fertile ground for further research. Be it data-driven applications and case studies which can benefit from the wealth of the available empirical data. Or perhaps a formal approach aimed at the method itself which might benefit from connections to an existing well-studied formalism, such as abstract interpretation~\cite{CousotC77}. Indeed, the construction of the generating rule set can be seen as an abstraction function, while the application of the rule set to the input graphs is the concretization function, together making up a Galois connection~\cite{CousotC92}.

%% file: manuscript.bbl
\begin{thebibliography}{10}
\providecommand{\url}[1]{\texttt{#1}}
\providecommand{\urlprefix}{URL }
\expandafter\ifx\csname urlstyle\endcsname\relax
  \providecommand{\doi}[1]{doi:\discretionary{}{}{}#1}\else
  \providecommand{\doi}{doi:\discretionary{}{}{}\begingroup
  \urlstyle{rm}\Url}\fi
\providecommand{\eprint}[2][]{\url{#2}}

\bibitem{AharoniGKGRT05}
Aharoni A, Gaidukov L, Khersonsky O, Gould SM, Roodveldt C, Tawfik DS.
\newblock The 'evolvability' of promiscuous protein functions.
\newblock \emph{Nature Genetics}, 2005.
\newblock \textbf{37}(1):73--76.
\newblock \doi{10.1038/ng1482}.

\bibitem{AndersenFMS14}
Andersen JL, Flamm C, Merkle D, Stadler PF.
\newblock Generic strategies for chemical space exploration.
\newblock \emph{International Journal of Computational Biology and Drug
  Design}, 2014.
\newblock \textbf{7}(2-3):225--258.
\newblock \doi{10.1504/IJCBDD.2014.061649}.

\bibitem{AndersenFMS16}
Andersen JL, Flamm C, Merkle D, Stadler PF.
\newblock {A software package for chemically inspired graph transformation}.
\newblock In: Echahed R, Minas M (eds.), Graph Transformation. ICGT 2016.
  Lecture Notes in Computer science, volume 9761, pp. 73--88. Springer, Cham,
  Switzerland.
\newblock ISBN 978-3-319-40529-2, 2016.
\newblock \doi{10.1007/978-3-319-40530-8_5}.

\bibitem{AndersenFMS19}
Andersen JL, Flamm C, Merkle D, Stadler PF.
\newblock Chemical Transformation Motifs—Modelling Pathways as Integer
  Hyperflows.
\newblock \emph{IEEE/ACM Transactions on Computational Biology and
  Bioinformatics}, 2019.
\newblock \textbf{16}(2):510--523.
\newblock \doi{10.1109/TCBB.2017.2781724}.

\bibitem{AppayeeB14}
Appayee C, Breslow R.
\newblock Deuterium Studies Reveal a New Mechanism for the Formose Reaction
  Involving Hydride Shifts.
\newblock \emph{Journal of the American Chemical Society}, 2014.
\newblock \textbf{136}(10):3720--3723.
\newblock \doi{10.1021/ja410886c}.

\bibitem{BehrK2021}
Behr N, Krivine J.
\newblock Compositionality of {R}ewriting {R}ules with {C}onditions.
\newblock \emph{{Compositionality}}, 2021.
\newblock \textbf{3}.
\newblock \doi{10.32408/compositionality-3-2}.

\bibitem{BennerKKR10}
Benner SA, Kim HJ, Kim MJ, Ricardo A.
\newblock Planetary Organic Chemistry and the Origins of Biomolecules.
\newblock \emph{Cold Spring Harbor Perspectives in Biology}, 2010.
\newblock \textbf{2}(7).
\newblock \doi{10.1101/cshperspect.a003467}.

\bibitem{BlinovFGH04}
Blinov ML, Faeder JR, Goldstein B, Hlavacek WS.
\newblock BioNetGen: software for rule-based modeling of signal transduction
  based on the interactions of molecular domains.
\newblock \emph{Bioinformatics}, 2004.
\newblock \textbf{20}(17):3289--3291.

\bibitem{BoutillierMLMKFCAF18}
Boutillier P, Maasha M, Li X, Medina-Abarca HF, Krivine J, Feret J, Cristescu
  I, Forbes AG, Fontana W.
\newblock The Kappa platform for rule-based modeling.
\newblock \emph{Bioinformatics}, 2018.
\newblock \textbf{34}(13):i583--i592.

\bibitem{BornscheuerHKLMR12}
Bornscheuer UT, Huisman GW, Kazlauskas RJ, Lutz S, Moore JC, Robins K.
\newblock Engineering the third wave of biocatalysis.
\newblock \emph{Nature}, 2012.
\newblock \textbf{485}(7397):185--194.
\newblock \doi{10.1038/nature11117}.

\bibitem{Breslow59}
Breslow R.
\newblock On the mechanism of the formose reaction.
\newblock \emph{Tetrahedron Letters}, 1959.
\newblock \textbf{1}(21):22--26.
\newblock \doi{https://doi.org/10.1016/S0040-4039(01)99487-0}.

\bibitem{BrusELP87}
Brus TH, van Eekelen MCJD, van Leer MO, Plasmeijer MJ.
\newblock {CLEAN:} {A} language for functional graph writing.
\newblock In: Kahn G (ed.), Functional Programming Languages and Computer
  Architecture, Portland, Oregon, USA, September 14-16, 1987, Proceedings,
  volume 274 of \emph{Lecture Notes in Computer Science}. Springer, 1987 pp.
  364--384.
\newblock \doi{10.1007/3-540-18317-5\_20}.

\bibitem{Butlerov61}
Butlerow A.
\newblock Formation synth{\'e}tique d’une substance sucr{\'e}e.
\newblock \emph{CR Acad. Sci}, 1861.
\newblock \textbf{53}:145--147.

\bibitem{ChaitinAC95}
Chaitin GJ, Arslanov A, Calude CS.
\newblock Program-size Complexity Computes the Halting Problem.
\newblock \emph{Bull. EATCS}, 1995.
\newblock \textbf{57}.

\bibitem{Cleaves11}
Cleaves HJJ.
\newblock Formose Reaction, pp. 600--605.
\newblock Springer Berlin Heidelberg, Berlin, Heidelberg.
\newblock ISBN 978-3-642-11274-4, 2011.
\newblock \doi{10.1007/978-3-642-11274-4_587}.

\bibitem{CopleyNW23}
Copley SD, Newton MS, Widney KA.
\newblock How to Recruit a Promiscuous Enzyme to Serve a New Function.
\newblock \emph{Biochemistry}, 2023.
\newblock \textbf{62}(2):300--308.
\newblock \doi{10.1021/acs.biochem.2c00249}.

\bibitem{CousotC77}
Cousot P, Cousot R.
\newblock Abstract Interpretation: A Unified Lattice Model for Static Analysis
  of Programs by Construction or Approximation of Fixpoints.
\newblock In: Proceedings of the 4th ACM SIGACT-SIGPLAN Symposium on Principles
  of Programming Languages, POPL '77. Association for Computing Machinery, New
  York, NY, USA.
\newblock ISBN 9781450373500, 1977 p. 238–252.
\newblock \doi{10.1145/512950.512973}.

\bibitem{CousotC92}
Cousot P, Cousot R.
\newblock Comparing the Galois Connection and Widening/Narrowing Approaches to
  Abstract Interpretation.
\newblock In: Proceedings of the 4th International Symposium on Programming
  Language Implementation and Logic Programming, PLILP '92. Springer-Verlag,
  Berlin, Heidelberg.
\newblock ISBN 3540558446, 1992 p. 269–295.

\bibitem{DelidovichSTP14}
Delidovich IV, Simonov AN, Taran OP, Parmon VN.
\newblock Catalytic Formation of Monosaccharides: From the Formose Reaction
  towards Selective Synthesis.
\newblock \emph{ChemSusChem}, 2014.
\newblock \textbf{7}(7):1833--1846.
\newblock \doi{https://doi.org/10.1002/cssc.201400040}.

\bibitem{EhrigEHP04}
Ehrig H, Ehrig K, Habel A, Pennemann KH.
\newblock Constraints and Application Conditions: From Graphs to High-Level
  Structures.
\newblock In: Ehrig H, Engels G, Parisi-Presicce F, Rozenberg G (eds.), Graph
  Transformations. Springer Berlin Heidelberg, Berlin, Heidelberg.
\newblock ISBN 978-3-540-30203-2, 2004 pp. 287--303.

\bibitem{EhrigPS73}
Ehrig H, Pfender M, Schneider HJ.
\newblock Graph-grammars: An algebraic approach.
\newblock In: 14th Annual Symposium on Switching and Automata Theory ({SWAT}
  1973). USA, 1973 pp. 167--180.
\newblock \doi{10.1109/SWAT.1973.11}.

\bibitem{HabelMP01}
Habel A, Müller J, Plump D.
\newblock Double-pushout graph transformation revisited.
\newblock \emph{Mathematical Structures in Computer Science}, 2001.
\newblock \textbf{11}(5):637–688.
\newblock \doi{10.1017/S0960129501003425}.

\bibitem{HeckelT20}
Heckel R, Taentzer G.
\newblock Graph transformation for software engineers.
\newblock Springer Nature, Cham, Switzerland, 1 edition, 2020.

\bibitem{Jensen76}
Jensen RA.
\newblock ENZYME RECRUITMENT IN EVOLUTION OF NEW FUNCTION.
\newblock \emph{Annual Review of Microbiology}, 1976.
\newblock \textbf{30}(1):409--425.
\newblock \doi{10.1146/annurev.mi.30.100176.002205}.

\bibitem{Karp72}
Karp RM.
\newblock Reducibility among Combinatorial Problems, pp. 85--103.
\newblock Springer US, Boston, MA.
\newblock ISBN 978-1-4684-2001-2, 1972.
\newblock \doi{10.1007/978-1-4684-2001-2_9}.

\bibitem{Kniemeyer08}
Kniemeyer O.
\newblock Design and implementation of a graph grammar based language for
  functional-structural plant modelling.
\newblock Ph.D. thesis, BTU Cottbus-Senftenberg, 2008.

\bibitem{Kolmogorov98}
Kolmogorov A.
\newblock On tables of random numbers.
\newblock \emph{Theoretical Computer Science}, 1998.
\newblock \textbf{207}(2):387--395.
\newblock \doi{https://doi.org/10.1016/S0304-3975(98)00075-9}.

\bibitem{LoboVD11}
Lobo D, Vico FJ, Dassow J.
\newblock Graph grammars with string-regulated rewriting.
\newblock \emph{Theoretical Computer Science}, 2011.
\newblock \textbf{412}(43):6101--6111.
\newblock \doi{https://doi.org/10.1016/j.tcs.2011.07.004}.

\bibitem{PallmannSHLHT18}
Pallmann S, \v{S}teflov\'{a}~(ne\'{e} Svobodov\'{a}) J, Haas M, Lamour S,
  Hen\ss A, Trapp O.
\newblock Schreibersite: an effective catalyst in the formose reaction network.
\newblock \emph{New Journal of Physics}, 2018.
\newblock \textbf{20}(5):055003.
\newblock \doi{10.1088/1367-2630/aabb99}.

\bibitem{RicardoCOB04}
Ricardo A, Carrigan MA, Olcott AN, Benner SA.
\newblock Borate Minerals Stabilize Ribose.
\newblock \emph{Science}, 2004.
\newblock \textbf{303}(5655):196--196.
\newblock \doi{10.1126/science.1092464}.

\bibitem{Rosen11}
Rosen K.
\newblock Discrete mathematics and its applications.
\newblock McGraw-Hill Professional, New York, NY, 7 edition, 2011.
\newblock ISBN 978-0-07-338309-5.

\bibitem{Socha80}
Socha RF, Weiss AH, Sakharov MM.
\newblock Autocatalysis in the formose reaction.
\newblock \emph{Reaction Kinetics and Catalysis Letters}, 1980.
\newblock \textbf{14}(2):119--128.
\newblock \doi{10.1007/BF02061275}.

\bibitem{Vazirani03_ch2}
Vazirani VV.
\newblock Set Cover, pp. 15--26.
\newblock Springer Berlin Heidelberg, Berlin, Heidelberg.
\newblock ISBN 978-3-662-04565-7, 2003.
\newblock \doi{10.1007/978-3-662-04565-7_2}.

\bibitem{Vazirani03_ch13}
Vazirani VV.
\newblock Set Cover via Dual Fitting, pp. 108--117.
\newblock Springer Berlin Heidelberg, Berlin, Heidelberg.
\newblock ISBN 978-3-662-04565-7, 2003.
\newblock \doi{10.1007/978-3-662-04565-7_13}.

\end{thebibliography}
